\documentclass[final,3p,twocolumn]{elsarticle}
\usepackage{latexsym}
\usepackage{amssymb, amsmath, bm, physics}
\usepackage{subcaption}
\usepackage{mathptmx}
\usepackage{aas_macros}
\usepackage{graphicx}

\journal{Physics of the Dark Universe}

\begin{document}

\begin{frontmatter}

\title{A quantum cosmology approach to cosmic coincidence and inflation}

\author[1]{S. Jalalzadeh}
\ead{shahram.jalalzadeh@ufpe.br}
\address[1]{Departamento de Fisica, Universidade Federal de Pernambuco, Recife, PE 50670-901, Brazil}
\author[2]{A. Mohammadi}
\ead{a.mohammadi@uok.ac.ir}
\address[2]{Department of Physics, Faculty of Science, University of Kurdistan,
Pasdaran Street, P.O. Box 66177-15175 Sanandaj, Iran}
\author[3]{D. Demir}
\ead{durmus.demir@sabanciuniv.edu}
\address[3]{Sabanc{\i} University, Faculty of Engineering and Natural Sciences, 34956 Tuzla, {\.I}stanbul, Turkey}

\begin{abstract}
This work studies the quantum cosmology of a closed, spatially homogeneous, and isotropic FLRW minisuperspace model with electromagnetic radiation as a  matter content. We solve the associated Wheeler--DeWitt (WDW) equation using the holographic regularization method and show that the electromagnetic zero-point energy forms the vacuum energy and provides a unified resolution to the horizon, flatness, singularity, and cosmic coincidence problems. This quantum cosmology approach composes an alternative to the usual inflationary paradigm and can be extended to more general matter contents and emergent gravity schemes.
\end{abstract}

\begin{keyword}
Quantum cosmology\sep Holography \sep Cosmic coincidence \sep UV/IR mixing \sep Inflation\sep Vacuum energy
\end{keyword}

\end{frontmatter}

\section{Introduction}
Observations since the mid-1980s have shown that the cosmological constant (CC) is small but nonzero \citep{Weinberg1988}. The Cosmic Microwave Background (CMB) observations point to a spatially flat universe, with $\Omega^{(\Lambda)}_0$ accounting for around 70\% of the energy density and non-relativistic (dark and baryonic) matter accounting for 30\% \citep{Planck2018}. Several dark energy candidates have been proposed in the literature, ranging from a small CC to exotic fields (quintessence, tachyon, k-essence, etc.) with appropriately determined potentials \citep{DE-review}. The majority of the candidates, however, suffer from the coincidence problem -- the question of why the matter (including the dark matter) and dark energy densities are of the same order today \citep{c-coincidence,Wang:1999fa,Dalal:2001dt,Griest:2002cu,Pavon:2005yx}. 

The CC, the simplest form of dark energy, appears to have contradictory roles from the field-theoretic and cosmological perspectives. Indeed, from the perspective of effective field theories (EFT), the CC is the zero-point energy set by some hard UV momentum cutoff such as the Planck scale, grand unification scale, or some other large scale beneath. From the perspective of cosmology, however, the same CC acts as the IR scale, setting the universe's large-scale structure. In this sense, the CC goes against UV/IR decoupling -- an indispensable feature for the naturalness of the EFTs. Indeed, the UV cutoff giving cause to the CC also motivates UV-sized masses for scalars and gauge fields so that there arises a strong UV/IR mixing \citep{UV-IR1,UV-IR2} together with an explicit breakdown of the gauge symmetries (including the electromagnetism) \citep{gauge-break1,gauge-break2,gauge-break3}. The EFT attains naturalness if  this UV/IR mixing and gauge breaking are prevented in a comprehensive framework. One such viable framework is  
symmergent gravity \citep{demir-2016,demir-2019,demir-2021}. (We keep in mind symmergent gravity as the solution, but  we do not explore it further as we focus mainly on electromagnetic radiation in the present work.) In essence, as it can be interpreted as both the zero-point energy and scale of the observable  universe, the CC goes beyond the local field-theoretic framework by blending the local UV and global IR physics \citep{holography-appl}. In fact, it is a possibility that the CC problem might be viewed as a quantum gravity problem, with the CC being a fundamental constant of nature in Einstein field equations \citep{Gurzadyan:2018dgm,2019EPJP98G,Gurzadyan:2021gek,Jalalzadeh:2021oxi}. (In the emergent gravity approach of the symmergent gravity, vacuum energy involves not the UV cutoff but the logarithmic part set by the particle masses  \citep{demir-2016,demir-2019,demir-2021}.)

The universe should have undergone a brief period of extremely rapid exponential expansion -- the cosmic inflation --  according to the standard model of cosmology  \citep{PhysRevD.23.347,Linde:1984ir}. Inflation can account for the large size, high entropy, and extreme spatial flatness of the current universe through a brief phase of accelerated expansion. The inflationary scenarios typically demand the existence of a scalar field (the inflaton) or more involved structures and add, therefore,  new variables to the hot big bang model \citep{Linde:1984ir}. The potential energy density of the scalar field has to adhere to various ``slow-roll'' conditions in order for it to be rolling for a long period of time. To enable energy transfer from the scalar field to matter at the end of inflation, it is also necessary to properly couple the scalar field to regular matter with appropriate initial conditions. The quantum fluctuations of matter generically disrupt the flatness of the inflaton potential, and it is necessary to keep  the induced density fluctuations below the observational bounds. The slope of the potential gets adjusted for slow roll after all these conditions are met. On top of all these, one notes that the big bang cosmology  fails to explain how the universe has arrived at its initial extremely hot and dense state. The question of what existed before the bang is a much more involved question for the standard cosmological paradigm. 

In this paper, we attempt to address the aforementioned problems of standard inflationary cosmology by constructing a simple quantum cosmological model. We construct a minimal quantum cosmology model in the presence of electromagnetic radiation (Sec. 2). Using the holographic principle \citep{holography1,holography2,holographyreview}, we show that the classical universe we know begins at a significantly large scale factor compared to the Planck length due to the existence of finite vacuum energy, to begin with (Sec. 3). As will be shown in the text, a classical universe can emerge without singularity, horizon, and flatness problems (Sec. 4). In addition, we put forth a new non-anthropic approach to the cosmic coincidence puzzle, which could enable elimination of fine-tuning involved (Sec. 4). We conclude the work by giving future prospects (Sec. 5). 

\section{Radiation in classical FLRW cosmology}
Let us consider a spatially-closed ($k=1$), homogeneous, and isotropic metric as the background minisuperspace. It is described by the line element
\begin{eqnarray}\label{1-1}
ds^2=-N^2(t)dt^2+a^2(t)h_{ij}dx^idx^j,
\end{eqnarray}
in which $a(t)$ is the scale factor, $N(t)$ is the lapse function, and $h_{ij}dx^idx^j=d\chi^2+\sin^2(\chi)d\Omega_{(2)}^2$ is the line element on the unit 3-sphere $\mathbb S^3$. We do not go into details of the background matter content and assume simply that it does not couple to the electromagnetic field up to a quadratic order. A typical background field would be a scalar \citep{HAWKING1984257}. (The assumption that the background fields do not couple to the electromagnetic field is a simplifying assumption for various reasons, one being the UV-sized mass for the photon field. Our analysis remains largely intact as long as a mechanism like symmergence is available to erase the loop-induced photon mass \citep{demir-2021}.) Including all the inhomogeneous degrees of freedom, the total action takes the form
\begin{equation}\label{RR}
    S=S_0+S^{(2)},
\end{equation}
in which $S_0$ is the background action and $S^{(2)}$ is the quadratic part of the inhomogeneties. In general, $S^{(2)}$ includes contributions from matter and gravitational perturbations. In the quadratic action $S^{(2)}$ what we are interested in is the electromagnetic field, which obtains the quadratic action  \citep{PhysRevD.38.478}
\begin{multline}\label{1-2}
%\begin{split}
   S_\text{Maxwell}=
    -\frac{1}{4}\int_{\mathcal M} \sqrt{-g}F_{\mu\nu}F^{\mu\nu}d^4x=\\
     \frac{1}{2}\displaystyle\int \Big\{\frac{a}{N}\Big(h^{ij}\dot A_i\dot A_j+
    2A_0~\partial_t(^{(3)}\nabla^k A_{k})-A_0~^{(3)}\nabla^2A_0\Big)\\
       -\frac{N}{a}A^k\Big(2A_k+
         ~^{(3)}\nabla_k\,^{(3)}\nabla^iA_i-~^{(3)}\nabla^2A_k\Big)\Big\}\sqrt{h}dtd^3x,
 %   \end{split}
\end{multline}
after using the  aforementioned assumption that the electromagnetic field does not couple to the background matter to quadratic order. In action (\ref{1-2}), $g$ is the determinant of the space-time metric $g_{\mu\nu}$, overdot denotes time derivative, $^{(3)}\nabla_k$ is induced 3-dimensional spatial covariant derivative,  spatial indices are raised and lowered by the metric $h_{ij}$ of the unit 3-sphere, and the electromagnetic field tensor $F_{\mu\nu}$ is given in terms of the 4-vector potential $A_\mu$ as
$F_{\mu\nu} =A_{\nu,\mu}-A_{\mu,\nu}$. 

Having fixed the electromagnetic part of $S^{(2)}$, we now turn to the action $S^0$ of the background fields. It is given by 
\begin{equation}\label{1-3}
\begin{split}
    S^0&=\frac{1}{16\pi G}\int_{\mathcal M}\sqrt{-g}(R-2\Lambda)d^4x-
    \frac{1}{8\pi G}\int_{\partial\mathcal M}\sqrt{h}Kd^3x\\
    &=\frac{3\pi}{4G}\int_{t_i}^{t_f}\Big(-\frac{a\dot a^2}{N}+Na-\frac{\Lambda}{3}a^3\Big)dt,
    \end{split}
\end{equation}
for the FLRW metric in (\ref{1-1}) with the cosmological constant $\Lambda$ and extrinsic curvature $K$.

In the background in (\ref{1-3}), the scalar electromagnetic potential $A_0(t,x^i)$ can be expanded into scalar hyperspherical harmonics $Y_{jlm}$ ($0\leq l\leq j$, $-l\leq m\leq l$) as
\begin{equation}\label{1-8}
A_0(t,x)=\sum_{j=0}^\infty\sum_{l=0}^j\sum_{m=-l}^{l}g_{jlm}(t)Y_{jlm}(x),    
\end{equation}
thanks to the $SO(4)$ rotation group (symmetry of $\mathbb S^3$).  In (\ref{1-8}),  $g_{jlm}(t)$ are functions of the time $t$. In addition, the spatial components $A_i(t,x^j)$ of the electromagnetic 4-potential can be Fourier expanded in terms of the vector hyperspherical harmonics, $Y^{jlm}_{(B)n}$, $(B=0,1,2)$ as
%\begin{equation}\label{1-9}
%    \begin{split}
%Y^{jlm}_{(0)i}&:=\frac{1}{\sqrt{j(j+2)}}~^{(3)}\nabla_iY^{jlm},
%\\
%Y^{jlm}_{(1)i}&:=\frac{1}{\sqrt{l(l+1)}}\varepsilon_i^{\,\,bc}~^{(3)}\nabla_bY^{jlm}\,^{(3)}\nabla_c\cos(\chi),
%\\
%Y^{jlm}_{(2)i}&:=\frac{1}{j+1}\varepsilon_i^{\,\,bc}~^{(3)}\nabla_bY^{jlm}_{(1)c}.
 %   \end{split}
%\end{equation}
%Then, the expansion of $A_k$ takes the form
\begin{eqnarray}\label{1-10}
A_k=\sum_{B=0}^2\sum_{j=j_\text{min}}^\infty\sum_{l=l_\text{min}}^j\sum_{m=-l}^l f^{jlm}_{(B)}(t)Y^{jlm}_{(B)k},
\end{eqnarray}
where the expansion coefficients $f^{jlm}_{(B)}(t)$ depend only on the time coordinate $t$. Note that all hyperspherical vector harmonics vanish for $j=0$. Also, $Y^{jlm}_{(1),i}$ and $Y^{jlm}_{(2),i}$ are not well-defined for $l=0$. Therefore, these hyperspherical harmonics are defined only for $j\geqslant0$, and $l\geqslant0$ for $B=0$ and $l\geqslant1$ for $B=1,2$. From now on, for a compact notation, we shall use $J:=\{jlm\}$ as a collective label for the indices $j$, $l$, and $m$. Now, replacing the Fourier expansions (\ref{1-8}) and (\ref{1-10}) in the Maxwell action (\ref{1-2}) 
leads to the simple action integral (see \cite{Jalalzadeh:2022bgz} for calculational details)
\begin{multline}\label{1-13}
S_\text{Maxwell}=\displaystyle\int dt\Big\{\frac{a}{2N}\Big(\sum_{B=1}^2\sum_J\dot f^2_{(B)J}-2\sum_{BJ}g_{(B)J}\dot f_{(B)J}
\\
+\displaystyle\sum_{BJ}j(j+1)g^2_{(B)J}\Big)-\frac{N}{2a}\sum_{B=1}^2\sum_Jf^2_{(B)J}\Big\},
\end{multline}
in which the momenta corresponding to $g_{(B)J}$ and $f_{(0)J}$ are obviously zero. In fact, they are Lagrange multiplies, which reflect the gauge freedom of the electromagnetic field in its canonical form. We fix these gauge freedoms by setting $
f_{(0)J}=g_{(B)J}=0$. In the above, equations (\ref{1-8}), (\ref{1-10}) and  $^{(3)}\nabla^iY^J_{(1,2)i}=0$ reveal that these gauge-fixing conditions are equivalent to the Coulomb-type (or radiation) gauge condition $A_0=0$ and $^{(3)}\nabla^iA_i=0$. As a result, 
the action (\ref{1-13}) for the electromagnetic field reduces to
\begin{equation}\label{1-16}
S_\text{Maxwell}=\displaystyle\sum_{B,J}%\sum_{j=1}^\infty\sum_{l=1}^j\sum_{m=-l}^l
\displaystyle\int dt\Big\{\frac{a}{ 2N}\dot f^2_{(B)J}
-\frac{ N}{2a}{(j+1)^2}f^2_{(B)J}\Big\},
\end{equation}
in a way containing mere physical variables. Note that in the above action $B=1,2$, $j$ runs from 1 to infinity, $l$ runs from 1 to $j$ and $m$ suns from $-l$ to $l$. This radiation action plus the gravitational part of the background action gives rise to us the Arnowitt--Deser--Misner (ADM) Lagrangian
\begin{multline}
    \label{1-17a}
    L_\text{ADM}=\frac{3\pi}{4G}\Big(-\frac{a\dot a^2}{N}+Na-\frac{\Lambda}{3}a^3\Big)+\\
    \frac{1}{2}\sum_{B,j,l,m}\Big(\frac{a}{{N}}\dot f^2_{(B)J}-\frac{N}{a}(j+1)^2f^2_{(B)J}\Big),
\end{multline}
where contributions of background fields (say, scalars) are all suppressed. This Lagrangian leads to the gauge field equations of motion 
\begin{eqnarray}\label{1-18}
\frac{d}{dt}\left(\frac{a\dot f_{(B)J}}{ N}\right)+\frac{N(j+1)^2}{a}f_{(B)J}=0,
\end{eqnarray}
with the solution
\begin{equation}\label{1-19}
    f_{(B)J}=D_{(B)J}\sin\Big( (j+1) \eta+\theta\Big),
\end{equation}
in which $D_{(B)J}$ and $\theta$ are the constants of integration, and the new time parameter $\eta$ is defined through the lapse function $d\eta=\frac{N}{a}dt$.

Besides the gauge field equations (\ref{1-18}), gravitational degrees of freedom ($a$ and $N$) obey the  Friedmann equations (in the comoving frame $N=1$)
\begin{equation}\label{1-20}
  \begin{split}
   &H^2=-\frac{1}{a^2}+\frac{8\pi G}{3}\rho^\text{(rad)}_0\left(\frac{a_0}{a}\right)^4+\frac{\Lambda}{3},\\
        &\frac{\ddot a}{a}=-\frac{8\pi G}{3}\rho^\text{(rad)}_0\left(\frac{a_0}{a}\right)^4+\frac{\Lambda}{3},
 \end{split}
\end{equation}
in which $H=\dot a/a$ is the Hubble parameter, and
\begin{equation}\label{1-21}
    \rho^\text{(rad)}_0 = \frac{1}{4\pi^2a_0^4}\sum_{(B)J}(j+1)^2D_{(B)J}^2,
\end{equation}
is the energy density of radiation at cosmic time $t_0$ at which the scale factor was $a_0$.

In our analysis, we have not considered inhomogeneous gravitational modes in the ADM Lagrangian (\ref{1-17a}). In that regard, space-time metric must also be expanded around the
 background metric in (\ref{1-1}) as
 \begin{multline}
     \label{RR1}
     ds^2=-\Big\{N(t)^2(1+C(x))^2+N_i(t,x)N^i(t,x)\Big\}dt^2+\\2N_i(t,x)dx^idt+a(t)^2\Big(h_{ij}+\epsilon_{ij}(x)\Big)dx^idx^j.
 \end{multline}
Here, perturbation of the lapse function $C(x)$ can be expanded in terms of the hyperspherical scalar harmonics, similar to the scalar electromagnetic potential in Eq. (\ref{1-8}). The shift vectors $N_i(t,x)$ can be Fourier-expanded in the same way as the spatial components of the electromagnetic 4-potential in (\ref{1-10}). Also, the perturbation of the spatial part of metric $\epsilon_{ij}$ can be expanded in terms of the scalar, vector, and tensor hyperspherical harmonics \cite{Lifshitz:1963ps,Halliwell:1984eu}. As demonstrated in Refs. \cite{Halliwell:1984eu,wada1986quantum}, (see also \cite{Bouhmadi-Lopez:2002tje,Kiefer:1987ft,Grishchuk:1993ds}), most of the terms appearing in the quadratic part of the action functional (\ref{RR}) can be eliminated in principle using diffeomorphism on the three-sphere and appropriate lapse and shift functions. The only terms that cannot be gauged away in these expansions are pure transverse traceless tensor perturbations that represent the gravitational waves. The resulting WDW equation for the tensor perturbations represents a set of harmonic oscillators  with  background-dependent frequencies \cite{wada1986quantum,Grishchuk:1989ss,Sidorov:1989qx}.  As Ref. \cite{Rosales:1996wj} showed,  the other of these quantum gravity corrections for the spectrum of gravitons is $\Lambda G$, and it would possibly lie on the
range of frequencies technically accessible in some future
experimental measurement of the CMB anisotropy. Since we are primarily interested in the behavior of quantum fluctuations of the electromagnetic field, explicit expression of $S^{(2)}_\text{gravity}$ (gravitational waves) is not necessary for investigating the behavior of the FLRW spacetime under the perturbations of the electromagnetic field.

In view of the various equations mentioned above, the energy-momentum tensor of the electromagnetic radiation, $T^{\mu\nu}$, is expected to be in the form of a perfect fluid. It is necessary to check this. In this regard, in the standard definition of the energy-momentum tensor, the Lagrangian density in (\ref{1-2}) leads to the following $T^{\mu\nu}$ components  
\begin{equation}
   \begin{split}
     T^{00}&=\frac{1}{4\pi^2a^4N^2}\left\{\frac{a^2}{N^2}h^{mn}\dot A_m\dot A_n+A_m(2A^m-~^{(3)}\nabla^2A_m)\right\},\\
       T^{0i}&=0,\\
      T^{mn}&=\frac{1}{2\pi^2a^4N^2}\dot A^m\dot A^n-\frac{1}{2\pi^2a^4}\Big\{ 2A^mA^n-A^{(m}~^{(3)}\nabla^2A^{n)}-\\&A^k~^{(3)}\nabla^{(i}~^{(3)}\nabla^{j)}A_k\Big\}+h^{mn}\Big\{\frac{1}{4\pi^2a^4N^2}\dot A_k\dot A^k-\\&\frac{1}{4\pi^2a^4}(2A_kA^k-A^k~^{(3)}\nabla^2A_k)\Big\}.
   \end{split}
\end{equation}
Obviously, $T_{ij}$ is not diagonal, and thus, it does not enjoy the same symmetries as the spatial part of the metric $h_{ij}$ in the line element (\ref{1-1}). This shows that, in the FLRW spacetime, $T_{\alpha\beta}$ does not take the perfect fluid form $T_{\alpha\beta}= (\rho+p)u_\alpha u_\beta+pg_{\alpha\beta}$. As a result, the geometrical symmetries of Einstein's tensor are not acknowledged and imposed on the matter sector. To address this issue at the classical level, we normally take integration over the 3-sphere to get the average value for the energy-momentum tensor \cite{1988GReGr201M}. The average value takes the form of the perfect fluid, with the equation of state $p=\frac{1}{3}\rho$. In this averaging sense, symmetries of the Einstein tensor agree with those of the perfect fluid \cite{1988GReGr201M}. In the present work, this critical averaging  was performed implicitly via the structure of the minisuperspace Lagrangian (\ref{1-17a}) and, as a result, the right-hand side of the Friedmann equations  (\ref{1-20}) stands for average characteristics of the radiation field as a perfect fluid \cite{Man}. Moreover, the Einstein field equations become $G_{\alpha\beta}=8\pi G\langle 0|T_{\alpha\beta}|0\rangle$ at the semi-classical level in which expectation value of the energy-momentum tensor takes the perfect fluid form above. (Intriguing discussions and extensions of the ideas above can be found in Refs. \cite{Bertolami:1990je,Bento:1992wy,Galtsov:1999bef,Moniz:1990hf,Bertolami:1991cf}.) One concern is why we do not start with $A_i=A_i(t)$ and $A_0=g(t)$ in order to ensure a good fit with FLRW symmetries from the beginning. In this case what happens is that  solution of the field equations for the matter part takes the form  $A_i(\eta)=D_i\sin(\sqrt{2}\eta)$  for a homogeneous and isotropic vector field, with $d\eta=\frac{N}{a}dt$ and $g(t)$ being arbitrary functions. This ensures that the aforementioned option leads to monochromatic radiation, which contradicts with the measurements of the CMB spectrum. Moreover, to explain the CMB black body radiation, what is needed is a quantized radiation field coupled to gravity and this could be classically defined by the action (\ref{1-16}), with the added feature that the matter part of the reduced ADM Lagrangian (\ref{1-17a}) enjoys the gravitational symmetries as a result of the shape of the original action functional (\ref{1-1}). For a Born--Infeld type radiation action, for instance,  our reduction no longer works. Indeed, Born--Infeld Lagrangian for radiation field is given by \cite{Galtsov:1999bef,Dyadichev:2001su,Moniz:2002rd,VargasMoniz:2003syv,VargasMoniz:2010upl}
\begin{multline}
    L_\text{BI}^\text{tr}\simeq\\\beta^2\tr\left(1-\sqrt{1+\frac{1}{2\beta^2}F_{\alpha\beta}F^{\alpha\beta}-\frac{1}{16\beta^2}(F_{\alpha\beta}\tilde F^{\alpha\beta})^2} \right),
\end{multline}
in which $\beta$ stands for the maximal field strength and $\tilde F_{\alpha\beta}$ is the dual of $F_{\alpha\beta}$. These models are inherently anisotropic (or inhomogeneous) as there is no way to generate a reduced Lagrangian due to the appearance of the gauge field strength $F_{\alpha\beta}$ inside the square root.

In the present epoch, the first  Friedmann equation in (\ref{1-20}) can be rewritten as
\begin{equation}
\label{densities}
    \Omega^\text{(rad)}_0+\Omega^\text{(cur)}_0+\Omega^{(\Lambda)}_0=1,
\end{equation}
in which $\Omega^\text{(rad)}_0=\frac{8\pi G\rho^\text{(rad)}_0}{3H_0^2}$, $\Omega^\text{(cur)}_0=-\frac{1}{a_0^2H_0^2}$, $\Omega^{(\Lambda)}_0=\frac{\Lambda}{3H_0^2}$ are, respectively, the density parameters for radiation,  curvature, and CC.
Let us go back to the cosmic coincidence problem. In this regard, let us consider, for a moment, adding dust (namely cold dark matter plus baryonic matter) to the Friedmann equations. The density parameter of dust, like the other matter components, is defined by $\Omega^\text{(dust)}_0=\frac{8\pi G\rho^\text{(dust)}_0}{3H_0^2}$ where $\rho^\text{(dust)}_0$ is the dust density at the present epoch. Existing observational data \citep{Planck2018} give $\Omega^\text{(dust)}_0\simeq\Omega^{(\Lambda)}_0$. This astounding fact that dark energy and dark matter assume similar energy density parameters now suggests that we are in a unique period in the whole of cosmic history. Thus, from the cosmic perspective, we happen to live in a brief period in which the matter and vacuum densities are of similar size \cite{Carroll:2000fy}. This coincidence has been dubbed as the ``coincidence problem'' \cite{c-coincidence,Wang:1999fa,Dalal:2001dt,Griest:2002cu,Pavon:2005yx}.  The observed values of the density parameters \citep{Planck2018}
 \begin{equation}\label{Omega1}
   \Omega^\text{(rad)}_0=8.7\times 10^{-5},~~~  \Omega^\text{(dust)}_0=0.315, ~~~\Omega^{(\Lambda)}_0=0.685,
 \end{equation}
suggest that not only do the matter and vacuum densities have comparable values within an order of magnitude but all density parameters do as well. In other words
\cite{Arkani-Hamed:2000ifx,Jamil:2012nma,Jamil:2011iu}
\begin{eqnarray}\label{Now}
    \frac{\Omega^\text{(rad)}_\text{now}}{\Omega^{(\Lambda)}_\text{now}}\simeq\frac{ \Omega^\text{(dust)}_\text{now}}{\Omega^{(\Lambda)}_\text{now}}\leq\mathcal O(1),
\end{eqnarray}
in a few orders of magnitude.
The occurrence of such a period is referred to as a ``triple coincidence'' or sometimes ``why now problem'' in the literature \cite{Arkani-Hamed:2000ifx,Jamil:2012nma,Jamil:2011iu}. It is worth noting that the ratios in the preceding relation are not equal at the present time. Indeed, regarding the present values of density parameters in (\ref{Omega1}), one finds
\begin{equation}
    \label{RRR}
    \frac{ \Omega^\text{(rad)}_0}{\Omega^{(\Lambda)}_0}=1.27\times10^{-4},~~~~~\frac{ \Omega^\text{(dust)}_0}{\Omega^{(\Lambda)}_0}=0.46\, .
\end{equation}
These hierarchically different ratios show that if the ``now" in the relation (\ref{Now}) stands for the present time, $t_0$, then the first fraction is off by a factor of three orders of magnitude. In consequence, a more precise definition of the moment ``now" or of the ``triple coincidence problem'' is needed. To this end, according to Ref. \cite{21W}, the problem is defined precisely if the ``now'' is defined as the period following matter-radiation equality in which the large-scale structure can arise. Then, the ``why now problem'' is rephrased as ``Why does the cosmological constant become comparable to other components of the cosmos right after matter-radiation equality\footnote{The critical scale in the matter power spectrum is the matter-radiation
equality scale (which defines the turn-around in the spectrum). The onset of
nonlinearity occurs when the density perturbation obeys $\delta\geq 3$. For a comoving scale
of 100 Mpc, this occurs at redshift $z\simeq1$.}?"  A more precise definition is given in Ref. \cite{Binetruy:2014yox} as follows: ``Why does dark energy become the dominant component at a period when galaxy formation is practically complete?"
\begin{figure}[h]
    \centering
    \includegraphics[width=8cm]{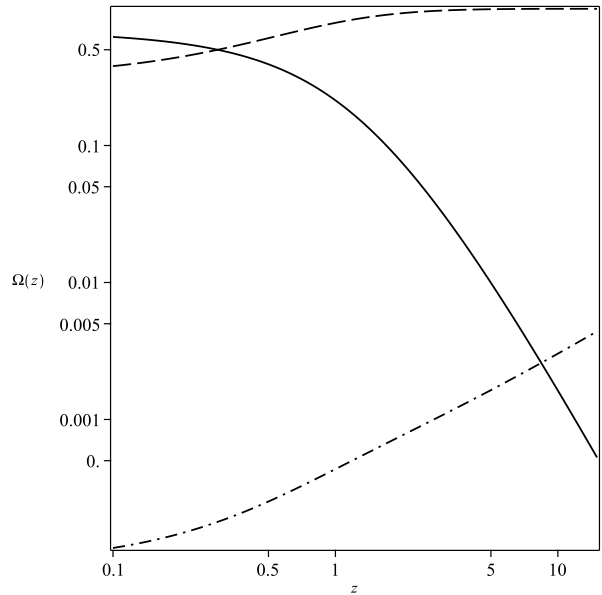}
    \caption{A log-log plot of the energy density parameter of dust, (dash), the radiation, (dash-dot), and the CC, (solid line) versus the redshift, $z$.}
    \label{fig0}
\end{figure}
In this sense, as revealed in Fig. \ref{fig0}, the ``now'' means a time period between redshifts $z=1$ and $z=3-5$ \cite{21W}.

\begin{figure}[h]
    \centering
    \includegraphics[width=8cm]{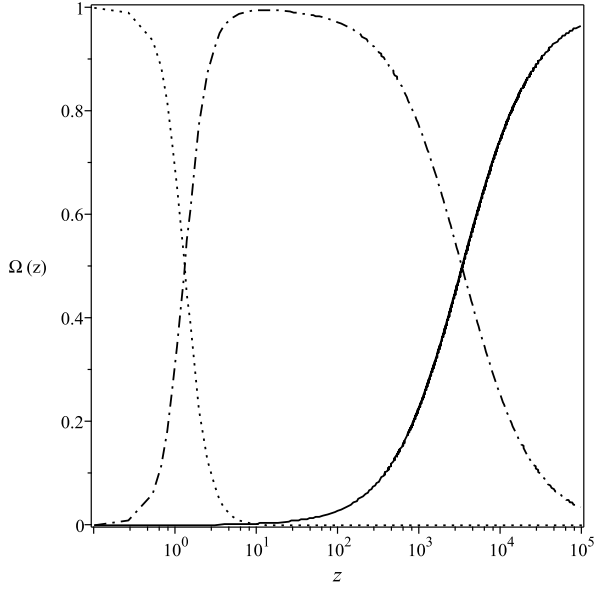}
    \caption{Evolution of  energy density parameter of the dust, $\Omega^\text{(dust)}(z)$, (dash-dot), the radiation, $\Omega^\text{(rad)}(z)$  (solid line), and the CC, $\Omega^{(\Lambda)}(z)$, (dot) as functions of the redshift, $z$.}
    \label{fig1}
\end{figure}

The three matter components' density parameters are depicted in Fig. \ref{fig1} as a function of redshift, $z$. The density parameters of radiation and dust are equal to the density parameter of CC at $z=08.9$ and $z=10^5$.
Note that in this figure the curvature part has been neglected. It demonstrates that a small portion of the redshift range is taken  by the dark-energy dominant period. This figure does, in fact, imply that we are in a highly unique phase of cosmic evolution.
This near-equality of the three physically distinct densities is highly perplexing. There is no question that the ``current'' epoch's equilibrium is highly advantageous for our survival. Indeed, for a closed model ($k=1$), harmony between $\Omega^\text{(rad)}$ and $\Omega^\text{(cur)}$ for a duration of the order of the Planck period, for example, would be catastrophic simply because, according to the preceding Friedmann equations, the entire period of existence would then be a few Planck times until re-collapse. For an open universe ($k=-1$), on the other hand, similar harmony would cause the universe to expand so quickly that, at the current epoch, it would be catastrophically small and incapable of enabling matter condensation. It is fortunate for these matters that the curvature and gravitational densities are almost equal at present.

 It is clear that, in the present work, we are just contemplating the radiation part of the coincidence problem. In this regard, let us turn to the ratios of the radiation density and the CC, given in equation (\ref{RRR}). We can address the coincidence problem by extrapolating the triple coincidence at the time of matter-radiation equality to today as follows: ``Why is the radiation density parameter some ten thousand times smaller than the dark energy density parameter at $t_0$?" The goal of the present work will be to realize/confirm this part of the triplet coincidence problem.

Corresponding to the ADM Lagrangian (\ref{1-17a}), the ADM Hamiltonian takes the form 
\begin{multline}
    \label{1-22}
    H_\text{ADM}=N\Big\{-\frac{G}{3\pi a}\Pi_a^2-\frac{3\pi}{4G}a+\frac{\pi\Lambda}{4G}a^3+\\
    \frac{1}{2a}\sum_{(B)J}\left(\Pi_{(B)J}^2+(j+1)^2f_{(B)J}^2\right) \Big\},
\end{multline}
in which $\Pi_a=-\frac{3\pi}{2G}\frac{a\dot a}{N}$ and $\Pi_{(B)J}=\frac{a}{N}\dot f_{(B)J}$ are the conjugate momenta for the scale factor $a$ and for the electromagnetic field $f_{(B)J}$, respectively. This ADM Hamiltonian leads us to the super-Hamiltonian constraint
\begin{multline}\label{1-23}
    \mathcal H=-\frac{G}{3\pi a}\Pi_a^2-\frac{3\pi}{4G}a+\frac{\pi\Lambda}{4G}a^3+   \\ \frac{1}{2a}\sum_{(B)J}\left(\Pi_{(B)J}^2+(j+1)^2f_{(B)J}^2\right)=0,
\end{multline}
which is what is expected of Hamiltonian gravity. (There is no external time in the sense of Schr\"odinger evolution.) Now, as preparation for quantization, let us introduce the complex-valued (annihilation and creation) variables
\begin{equation}
   \begin{split}\label{1-24}
   C_{(B)J}&=\frac{1}{\sqrt{2(j+1)}}\left(i\Pi_{(B)J}+(j+1)f_{(B)J}\right),\\ 
   C^*_{(B)J}&=\frac{1}{\sqrt{2(j+1)}}\left(-i\Pi_{(B)J}+(j+1)f_{(B)J}\right),
\end{split} 
\end{equation}
so that the set $S=\{C_{(B)J},C^*_{(B)J},1 \}$ is closed under the Poisson bracket $\{C^*_{(B)J},C_{(B')J'}\}=i\delta_{BB'}\delta_{JJ'}$. It is clear that every sufficiently differentiable function in the phase space of the matter sector can be expressed in terms of $S$. Now, using the conjugate variables in (\ref{1-24}) the ADM Hamiltonian in (\ref{1-22}) can be recast as \begin{multline}
    \label{1-25}
 H_\text{ADM}=N\Big\{-\frac{G}{3\pi a}\Pi_a^2-\frac{3\pi}{4G}a+\frac{\pi\Lambda}{4G}a^3+\\
    \frac{1}{a}\sum_{(B)J}(j+1)C^*_{(B)J}C_{(B)J} \Big\},
\end{multline}
with respect to which the complex-valued variables evolve as 
\begin{equation}
%    \begin{split}
       C_{(B)J}=C_{(B)J}(0)e^{-i(j+1)\eta},~~
        C_{(B)J}^*=C^*_{(B)J}(0)e^{i(j+1)\eta},
 %   \end{split}
\end{equation}
as expected of (annihilation and creation type) Fourier decomposition coefficients. 
%These solutions lead us to the mode-expansion of the gauge fields  (\ref{1-10})
%\begin{equation}\label{1-27} \begin{split} A_i(\hat{x},\eta)=\sum_{(B)J}\frac{1}{\sqrt{2(j+1)}}\Big\{C_{(B)J}Y_{(B)i}^Je^{-i(j+1)\eta}+\\ C^*_{(B)J}Y_{(B)i}^{J*}e^{i(j+1)\eta}\Big\}.\end{split}\end{equation}
%The conjugate momenta of $A_i$ is $P^i=\delta S_m/\delta\dot A_i=\frac{ah^{ij}}{N}\dot A_j$ and the equal-time Poisson bracket is set to be $\{A_i(\eta,\hat{x}),P_j(\eta,\hat{ y})\}=\delta_{ij}(\hat{ x},\hat{{y}})$, where
%\begin{eqnarray}\label{1-28}\delta_{ij}(\hat{ x},\hat{{y}})=\sum_{(B)J}Y^{i*}_{(B)J}(\hat{{x}})Y^{j}_{(B)J}(\hat{{y}}),
%\end{eqnarray}
%is the delta function on $\mathbb S^3$.

\section{Quantum cosmology with electromagnetic radiation}

In order to obtain the WDW equation, we apply the usual first-quantization rules
%\begin{equation}\label{quan}
%\begin{split}
$(a, \Pi_a)\rightarrow (a,-i\frac{\partial}{\partial a})$, $(f_{(B)J},\Pi_{(B)J})\rightarrow (f_{(B)J},-i\frac{\partial}{\partial {f_{(B)J}}})$
%\end{split}
%\end{equation}
where vanishing of the Hamiltonian as in (\ref{1-23}) ensures time-independence of the dynamical system. In addition to these, we use the well-known Hartle--Hawking--Verlinde \citep{PhysRevD.28.2960,PhysRevD.33.3560,PhysRevD.37.888} operator ordering \footnote{There exist also two-parameter families of orderings in the literature  \citep{doi:10.1142/8540,Pedram:2008sj,Steigl:2005fk}. It is easy to demonstrate that the general discussion in this letter can be applied to various orderings, and the outcomes are identical to those found here.}
\begin{equation}
    \label{2-new1}
    \frac{1}{a}\Pi_a^2=a^{q-1}\Pi_aa^{-q}\Pi_a=-\frac{1}{a}\partial_a^2+\frac{q}{a^2}\partial_a,
\end{equation}
where $q$ is the factor ordering parameter. In consequence,  the application of the quantization map to the Hamiltonian constraint in (\ref{1-23}) leads to the WDW equation 
\begin{multline}
    \label{2-29}
    \Bigg\{\frac{1}{3\pi a M_P^2}\left(-\partial_a^2+\frac{q}{a}\partial_a\right)+\frac{3\pi M_P^2}{4}a-\\\frac{\pi\Lambda M_P^2}{4}a^3-{\mathcal H_\text{rad}}\Bigg\}\Psi(a,f_{(B)J})=0,
\end{multline}
after using the operator ordering in (\ref{2-new1}) with the Planck scale $M_P=1/\sqrt{G}=1/L_P$. The radiation part of the super-Hamiltonian 
\begin{equation}\label{2-31}
    \mathcal H_\text{rad}=\frac{1}{2a}\sum_{(B)J}\left(-\partial_{(B)J}^2+(j+1)^2f_{(B)J}^2\right),
\end{equation}
in the above WDW equation represents the contribution of an infinite number of harmonic oscillators (photons). The eigenfrequencies of the
electromagnetic field are $\omega_j=(j+1)/a$. Now, passing from $(f_{(B)J},\Pi_{(B)J})$ to the creation-annihilation operators $(C^\dagger_{(B)J}, C_{(B)J})$ with the commutation relation 
%\begin{eqnarray}\label{2-32}
    $\left[C_{(B)J},C^\dagger_{(B')J'}\right]=i\delta_{BB'}\delta_{JJ'}$.
%\end{eqnarray}
one can rewrite the radiation Hamiltonian (\ref{2-31}) as
\begin{equation}\label{2-33}
    \mathcal H_\text{rad}=\frac{1}{a}\sum_{(B)J}(j+1)\left(N_{(B)J}+\frac{1}{2}\right),
    %=\\\frac{1}{a}\displaystyle\sum_{B=1}^2\sum_{j=1}^\infty\sum_{l=1}^j\sum_{m=-l}^l(j+1)\left(N_{(B)jlm}+\frac{1}{2}\right),
\end{equation}
with the usual number operator $N_{(B)J} = C^\dagger_{(B)J}C_{(B)J}$. Then, the vacuum energy is found to be
\begin{equation}\label{2-34}
    \mathcal E_\text{rad}=\frac{1}{a}\sum_{j=1}^\infty j(j^2-1).
\end{equation}
It is clear that this is the total vacuum energy, deriving from the electromagnetic field in (\ref{1-8}) and (\ref{1-10}), induced by the zero-point energies of the individual photons. It is divergent (infinite). (If the electromagnetic field were interacting with a background field ${\mathcal B}$ as, say, ${\mathcal B}^\dagger {\mathcal B} A_\mu A^\mu$ then the field ${\mathcal B}$ would acquire a divergent mass from these zero-point fluctuations.)    One way to regularize the vacuum energy (\ref{2-34}) is to use the Riemann zeta function regularization method \citep{Jalalzadeh:2022bgz}. To do so, we must assume that elementary particles, or the electromagnetic field in our simplified model, can be accurately described by an EFT with a UV cutoff $M_{\Lambda}=\sqrt{\Lambda/3}$ (set by the CC $\Lambda$) below the Planck mass ($M_{\Lambda}< M_P$) so that the sum in (\ref{2-33}) stops at $j_{max}=M_{\Lambda}/M_P$, leading this way to finite vacuum energy. This EFT description works so long as all momenta and field strengths are small in comparison $M_{\Lambda}$ to appropriate power \citep{holography-appl}. Even if all these conditions are met, the EFT description gets hindered by the fact that energy density in a given medium is bounded by the critical density required by black hole formation. (In other words, the EFT cannot be put in a volume smaller than the horizon size). This constraint causes an unavoidable UV/IR mixing in that the entropy $S$ of a local field-theoretic system must obey the Bekenstein bound \citep{Bekenstein:1980jp}
\begin{equation}\label{Bek}
    S\leq 2\pi ER,
\end{equation}
pertaining to a self-gravitating physical
system of radius $R$ and energy $E$. This holographic bound \citep{holography1,holography2,holographyreview} relates maximal degrees of freedom in a given volume to its boundary surface area. The holographic principle has caused profound changes in our understanding of gravity. Its basic implication is that Lagrangian field theories substantially overcount the degrees of freedom so that the Bekenstein bound in (\ref{Bek}) eliminates those EFT states falling inside the black hole horizon \citep{holography1,holography2}. Indeed, Lagrangian field theories are characterized by entropic extensivity.  In a given EFT, the Bekenstein  bound (\ref{Bek}) can be satisfied by restricting the volume containing the quantum fields as $R^3 M_\Lambda^3 \leq S_{BH}$, where $S_{BH}$ is the entropy of a black hole of radius $R$. As a result, the radius $R$, which serves as the IR cutoff, scales as $1/M_\Lambda^{3}$ and cannot, therefore, be selected independently of the UV cutoff (as well as the CC $\Lambda$). Due to the restriction imposed by the formation of the black hole at high energy density, it has been hypothesized that in effective quantum field theory, a short-distance UV cutoff ($M_\Lambda$) is tied to the long-distance IR cutoff ($R$) \citep{holography-appl}. In specific terms,  if $\rho$ is the quantum zero-point energy density set by the short distance cutoff $M_\Lambda$ then the total energy in a region of size $R$ should not exceed the mass of the black hole of the same size. In fact, the largest possible $R$ (IR cutoff) is the one that saturates the inequality. In consequence, the electromagnetic vacuum energy (\ref{2-33}) satisfies the relation
\begin{equation}
    \label{bek2}
 4\pi G   \mathcal E^2_\text{rad}\Big|_{a=a_0}=S_{dS}=\frac{A_{dS}}{4G},
\end{equation}
where $L_\Lambda=\sqrt{3/\Lambda}=16.4\pm 0.4 ~\text{Glyr}= (1.55\pm 0.04)\times10^{26}~ \text{m}$ is de Sitter radius of the Universe, $A_{dS}=4\pi L_\Lambda^2$  is the area of the de Sitter horizon, and
\begin{equation}\label{Lambda}
    S_{dS}=\frac{A_{dS}}{4G}=2.88\times10^{122},
\end{equation}
is de Sitter entropy \citep{2010ApJ1825E}.

{One notes that $\mathcal E_\text{rad}$ in (\ref{bek2}) represents the total energy of the radiation in the present epoch, for which $a_0\simeq L_\Lambda$. Regarding Eq.(\ref{1-21}), one can estimate the energy density of CMB radiation at the present epoch by 
\begin{equation}
    \rho_0^\text{(rad)}=\frac{\mathcal E^{\text{(rad)}}_0}{4\pi^2a_0^3}\simeq\frac{S_\text{dS}^\frac{1}{2}}{8\pi^\frac{5}{2}L_PL_\Lambda^3}=\frac{1}{3\pi}\frac{\Lambda}{8\pi G}\simeq 10^{-1}\rho_0^{(\Lambda)},
\end{equation}
which is close to what is observed. In the last section, we will give a more accurate estimate of this energy density.}

Now, from (\ref{bek2}) one gets the relation
\begin{equation}
\label{Erad-sum}
    4\pi\frac{L_P^2}{L_\Lambda^2}\sum_{j=1}^{j_{max}} j(j^2-1)=4\pi\frac{L_\Lambda^2}{L_P^2},
\end{equation}
which is approximately saturated by $j_{max}$ as
\begin{equation}\label{abb0}
 \sum_{j=1}^{j_{max}} j(j^2-1)\simeq\frac{j_{max}^4}{4}=\frac{S_{dS}}{4\pi},
\end{equation}
so that one gets  $j_{max}=(S_{dS}/\pi)^{1/4}$ as the cutoff on the wave number of the virtual photons. One also finds $\mathcal E_\text{rad}=S_{dS}/4\pi a$ as the relation between the photon energy and the de Sitter entropy. In addition to these,  in a comoving volume $V$ and in the frequency band from $\nu=(j+1)/a$ to $\nu+\delta\nu=(j+2)/a$ the total number of virtual photons  is given by
\begin{equation}
    \delta N=8\pi V\nu^2\delta\nu=\frac{8\pi}{3}(j+1)^2.
\end{equation}
The use of this virtual photon number leads to the radiation entropy
\begin{equation}\label{abb1}
  S_\text{rad}= N=\frac{8\pi}{3}\sum_{j=1}^{j_{max}}(j+1)^2\simeq \frac{8\pi}{9}j_{max}^3=\frac{8\pi^\frac{1}{4}}{9}(S_{dS})^\frac{3}{4},
\end{equation}
where use has been made of the relation  (\ref{abb0}) in  the last equality.

Let us go back to the WDW equation  in (\ref{2-29}). It simplifies to 
\begin{equation}\label{2-39}
    -\psi''(a)+\frac{q}{a}\psi'(a)+\left(\frac{3\pi}{2L_P^2}\right)^2a^2\left(1-\frac{a^2}{L_\Lambda^2}\right)\psi(a)=\frac{L_\Lambda^2}{L_P^4}\psi(a),
\end{equation}
after taking the radiation energy from the holographic relation in (\ref{Erad-sum}). Here, prime on the wavefunction $\psi(a)$ denotes the derivative with respect to the scale factor $a$. The operator in the left-hand side of this WDW equation is defined on a dense domain $C^\infty(0,+\infty)$, and the necessary and sufficient condition for it to be a Hermitian operator is
$\frac{\psi(a)}{\psi'(a)}\Big|_{a=0}={\gamma}$,
where $\gamma$ is a real parameter. In essence, the parameter $\gamma$ defines a 1-parameter family of self-adjoint extensions of the WDW operator on the half-line. One notes that the last term in the left-hand side of (\ref{2-39}) can be treated as a perturbation as it is negligibly small for $a\ll L_\Lambda= 10^{30}L_P$. In fact, it could be significant only for $a\simeq L_\Lambda$.
Ignoring that term, % changing the minisuperspace variable $a$ into $\xi=\frac{3\pi m_P^2}{2}a^2$, and rewriting the wavefunction as $\psi=\exp(-\frac{\xi}{2})g(\xi)$ in Eq.(\ref{2-39})
the WDW equation is solved by the confluent hypergeometric function
\begin{multline}
    \label{new2-2}
    \psi(a)=\sqrt{2}\left(\frac{3\pi}{4L_p^4}\right)^\frac{q+3}{4}e^{-\frac{3\pi a^2}{4L_P^2}}a^{q+1}\cdot\\
    _1F_1\left(-\frac{1}{6\pi}\left( \frac{L_\Lambda}{L_P}\right)^2;\frac{q+3}{2};\frac{3\pi a^2}{2L_P^2}\right).
\end{multline}
%where $\lambda=\frac{6\pi^2}{G\Lambda}=2\pi^2(\frac{L_\Lambda}{L_P})^2$. This equation is the well known Kummer’s differential equation \cite{Abramowitz}, whose regular solution at the Big-Bang singularity, $\xi=0$, is the confluent hypergeometric function 
In order for the wavefunction to be square-integrable, the hypergeometric series $_1F_1(\alpha;\beta;\xi)$ must terminate, and this is achieved  if there exists some non-negative integer $n$ such that $\alpha=-n$. In consequence, one arrives at the quantization 
\begin{eqnarray}
    \label{new2-3}
 -\frac{1}{6\pi}\left( \frac{L_\Lambda}{L_P}\right)^2=- n~~~\text{implying}~~~~\frac{S_\text{dS}}{24\pi^2}=n.
\end{eqnarray}
This result comes to mean that, in accordance with the holographic principle, the entropy of the de Sitter space is equal to the number $n$ of fundamental cells  in the holographic screen \citep{doi:10.1142/8540,Jalalzadeh:2017jdo,Sheikh-Jabbari:2006nsl}. In consequence, the wavefunction $\psi(a)$ of the superspace becomes square-integrable automatically provided that the quantization condition (\ref{new2-3}) holds. The appearance of the fundamental cells rests on the assumption that the holographic principle is a fundamental principle of Nature (at the same level as the equivalence and relativity principles).

{One of the strangest and most enigmatic phenomena in general relativity is the existence of singularities, particularly cosmic singularities. A singularity is a point in the history of the universe where the conditions are so severe that none of the established physical laws hold. A fundamental problem that still lacks a clear answer concerns the irreversible collapse of the spacetime continuum. To this end, quantum effects have been proposed as a remedy for singularities.  As a concrete formulation, DeWitt proposed in Ref.\cite{PhysRev.160.1113} that the wavefunction of the universe must vanish at the classical singularity of the classical cosmological model considered. This criterion, known as the DeWitt boundary condition, was the first to eliminate the classical singularity. Furthermore, several authors have proposed extending DeWitt's idea to ensure conformal invariance(see, for example, \cite{Kiefer:2019bxk} and references therein). Nonetheless, as proved in Ref.\cite{PhysRevD.28.2402}, the DeWitt boundary condition has nothing to do with avoiding the singularities although its fundamental relevance has long been debated. In order to assess the quantum status of the singularity, one can consider evaluating the expectation values of observables which classically vanish at the singularity \cite{PhysRevD.28.2402,PhysRevD.22.235,PhysRevD.8.3253}. In consequence, a quantum state $|\psi\rangle$ is said to be singular if and only if every operator $Q$,  corresponding to a classical observable which vanishes at the singularity, has a vanishing vacuum expectation value 
$\langle\psi|Q|\psi\rangle=0$ at the singularity. This quantum collapse test is just as convincing as any other, and it has the added benefit of being very simple to implement and verify. In our model, the observable is $g_{rr}=a(t)^2$ -- the scale factor squared. To implement this condition in our model, we first rewrite the wavefunction (\ref{new2-2}) in terms of the generalized Laguerre polynomials, $L_n^{(\beta)}(x)$. When the first argument of the  confluent hypergeometric function is a negative integer, which is what happens for the wavefunction $\psi(a)$ in (\ref{new2-2}), we get the identity}
\begin{equation}
    \begin{pmatrix}
    n+\beta\\
    n
    \end{pmatrix}~_1F_1(-n;\beta+1; x)=L_n^{(\beta)}(x)
\end{equation}
{as a definition of the Laguerre polynomials. Now, the integral identity with  $x^\beta e^{-x}$}
\begin{equation}
\int_0^\infty x^\beta e^{-x}\left[L_n^{(\alpha)}(x)\right]^2dx=\frac{\Gamma(n+\beta+1)}{\Gamma(n+1)}
\end{equation}
{enables one to determine the normalization constant of the wavefunction (\ref{new2-2}). The integral identity with $x^{\beta+1} e^{-x}$}
\begin{equation}
         \int_0^\infty x^{\beta+1} e^{-x}\left[L_n^{(\alpha)}(x)\right]^2dx=\frac{\Gamma(n+\beta+1)}{\Gamma(n+1)}(2n+\beta+1),
\end{equation}
{on the other hand, leads to a determination of the expectation value of $g_{rr}$ as}
\begin{equation}
    \label{Omega3}
    \langle a^2\rangle=\frac{S_\text{dS}L_P^2}{6\pi^3}=\frac{L_\Lambda^2}{6\pi^2}
\end{equation}
{after using the definition of $S_\text{dS}$ in (\ref{Lambda}). This result shows that quantum singularity is no longer present. In addition, using the relation $_1F_1(0;\alpha;\beta)=1$ one can easily verify that the wavefunction (\ref{new2-2}) satisfies the DeWitt condition. This gives credence to our conclusion above.}

{The expectation value of the scale factor in (\ref{Omega3}) reveals an important result: the scale factor's expectation value is nearly equal to the size of the universe in the current epoch. One notes that the expectation value would be proportional to the Planck length in the case of cutoff regularization of the quantum fluctuations  \cite{Jalalzadeh:2022bgz}. And as a result, the quantum universe would recollapse back to the singularity in the Planck epoch \cite{PhysRev.160.1113}. In this case, an inflationary scalar field would be required to avoid the re-collapse of the nucleated universe. Our result in (\ref{Omega3}), however,  is a direct consequence of the holographic renormalization of the vacuum energy and entirely eliminates any need for an inflaton field. This is the main virtue of the  holographic renormalization of vacuum energy. As we will discuss in detail in the next section, holographically renormalized vacuum energy renders the inflationary fields unnecessary even for addressing  the usual problems with the standard hot Big Bang model.}

\section{Emergent classical universe}
{
The wavefunction in (\ref{new2-2}) is orthonormal for $(1+q)/2<1$. Moreover, the integer $n$, defined in (\ref{new2-3}), is nearly equal to the de Sitter entropy or the event horizon entropy of the Universe ($n\simeq10^{122}$). }
Therefore, one can use the 
asymptotic forms of the confluent hypergeometric functions \citep{Abramowitz} to arrive at the compact wavefunction
\begin{equation}
    \label{wa111}
    \psi(a)=\mathcal Ca^\frac{q+1}{2}\cos\left(\frac{L_\Lambda a}{2L_P^2}-\frac{\pi}{4}(q+2) \right).
\end{equation}
{One notes that this wavefunction proves a good approximation to the exact form of the wavefunction even at the Planck scale.  This is due to the tantalizingly large value of $n$. }
The oscillatory behavior of the above wavefunction reveals that the resulting universe is flat and expands without limit. In addition, as a result of the huge value of $L_\Lambda/L_P$, it oscillates very rapidly as  $\cos\left(6\pi n (a/L_{\Lambda})-\pi (q+2)/4\right)$ to put the universe into the semi-classical regime immediately after the Planck time. Indeed, it collects both the expanding $\exp(-iS)$ and contracting $\exp(iS)$, $S=\frac{L_\Lambda a}{2L_P^2}$, phases of the model universe simultaneously. In fact, for an expanding universe, {using the definition of the conjugate momenta of the scale factor  $\Pi_a=-\frac{3\pi}{2G}\frac{a\dot a}{N}$ in (\ref{1-22})}, one obtains
\begin{equation}\label{wave1}
    \Pi_a=-\frac{dS}{da}=-\frac{L_\Lambda}{2L_P^2},~~~\text{and}~~~\Pi_a=-\frac{3\pi}{2L_P^2}a\dot a,
\end{equation}
leading to the Hubble parameter
\begin{equation}\label{wave1aa}
    H =\frac{L_\Lambda}{{3}\pi}\frac{1}{a^2},
\end{equation}
as follows via the definition of $\Pi_a$ in equation (\ref{1-22}). This Hubble parameter gives rise to the scale factor 
\begin{equation}
    \label{wave2}
    a(t)=\sqrt{\frac{2 L_\Lambda t}{3\pi}},
\end{equation}
for $t\geq t_P$. 
{It is worth noting that this radiation-induced scale factor is not an extrapolation of the radiation-dominated classical Friedmann equations. It stands for the classical radiation-dominated classical universe emerging after the Planck time through the quantum cosmological effects.}
This solution is characteristic of the   radiation-dominated flat universe, with the only difference being that it is rescaled with a huge prefactor proportional to $\sqrt{L_\Lambda}$. For example, for the Planck time $t=t_P=L_P$ and in the GUT comoving time $t=t_G\simeq 10^{-36}~s$ the scale factor takes the respective values
\begin{equation}\label{wave3}
%  \begin{split}
      a(t_P)\simeq30~\mu{\rm m}~~~~~{\rm and}~~~~~ a(t_G)\simeq14~{\rm cm}.
%  \end{split}
\end{equation}
It is clear that the value of scale factor $a(t_P)$ at the beginning of time is given by the entropy of the holographic screen
\begin{equation}\label{wave3a}
    a(t_P)=\sqrt{\frac{2}{3\pi^\frac{5}{2}}}S_{dS}^\frac{1}{4}L_P\simeq 10^{30}L_P.
\end{equation}
{It is worth noting that the whole region with the aforementioned scale factor has arisen from a single wavefunction (\ref{wa111}) during the Planck period. As a result, the original state has had the huge homogeneous region above.}

As the universe gets bigger, that is, when $a\gg L_\Lambda$ the WDW equation (\ref{2-39}) acquires the solution
\begin{equation}\label{wave4}
\psi(a)=\mathcal C' a^\frac{(q-2)}{2} \cos\left(\frac{\pi a^3}{2L_P^2L_\Lambda}+\frac{(q-2)\pi}{12}\right),
\end{equation}
and its expanding universe solution leads to the semi-classical de Sitter space
\begin{equation}\label{wave5}
    \Pi_a=\frac{d}{da}\left(-\frac{\pi a^3}{2L_P^2L_\Lambda}\right)=-\frac{3\pi}{2L_P^2}a\dot a,
\end{equation}
with the Hubble constant 
\begin{equation}\label{wave6}
   H =\sqrt{\frac{\Lambda}{3}}.
\end{equation}
This Hubble parameter leads to the late time de Sitter expansion.

\begin{figure}[h!]
    \centering
    \subfloat[\centering]{{\includegraphics[width=7cm]{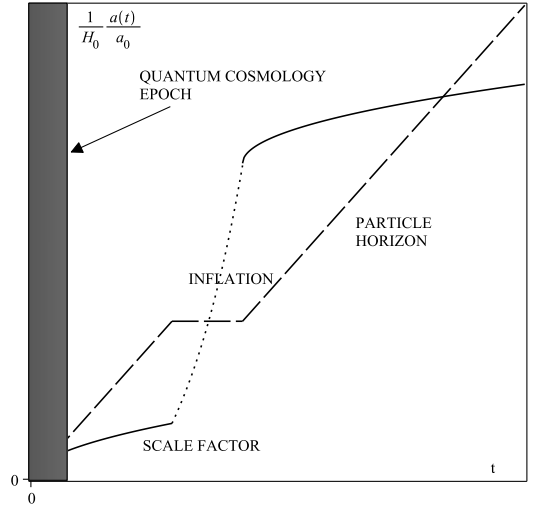} }}%
    \qquad
    \subfloat[\centering]{{\includegraphics[width=7cm]{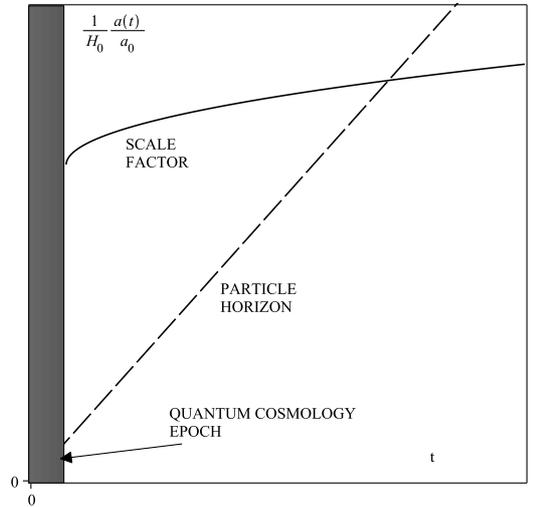} }}%
    \caption{{Evolution of the universe in the inflationary (panel (a)) and quantum cosmological (panel (b)) approaches. As shown in panel (a), inflation not only protects the universe from collapsing but also resolves the problems of the hot Big Bang. As seen from panel (b), on the other hand, once the nucleated quantum universe  enters the classical epoch, it is  large enough not to collapse back on itself. As panel (b) makes it clear quantum cosmological evolution brings the universe's state to the post-inflation epoch in panel (a), with no need for inflation.}}%
    \label{fig2-3}%
\end{figure}

In view of the quantum cosmological holographic solution above, we now discuss the basic problems of the hot big bang cosmology. Let us start with the horizon problem. It is well known that inflation is a brief phase of accelerated expansion. (It is essentially a hypothetical phase since the inflaton itself disappears at the end of the inflationary phase.) This phase should have most likely occurred  during the first few picoseconds when the size of the universe increased by a factor $10^{30}$ (some 60 e-foldings). As is seen from the relation (\ref{wave3}), when the universe interns into the classical realm its causally-connected scale is $10^{30}$ times the Planck length, and this patch size is enough to solve the horizon problem of the big bang cosmology, {see Fig. \ref{fig2-3} (a). After the universe enters the classical era, in the inflationary epoch, the scale factor's behavior changes exponentially and prevents the universe from re-collapsing. Furthermore, the particle horizon expands significantly, and the entire observable universe at the present epoch gets causally connected.}

Let us now discuss how the flatness problem is resolved by the adiabatic expansion of the universe. According to the standard cosmology, the Friedmann equations ensure that during the radiation dominance, the Hubble parameter is given by 
\begin{equation}
    \label{En1}
    H^2\simeq \frac{8\pi}{M_P^2}\rho_\text{rad}\simeq \frac{T^4}{M_P^2},
\end{equation}
where $T$ is the temperature ($\Lambda$ is completely negligible in this early phase of the cosmological evolution). Then, as a measure of the flatness \citep{Boerner:2003qv}
\begin{equation}
    \label{En2}
    \frac{|1-\Omega|}{\Omega}\Big|_{t=t_P}\simeq|k|S_\text{rad}^{-\frac{2}{3}}\left( \frac{M_P}{T}\right)^2\Big|_{T=T_P}\simeq 10^{-60},
\end{equation}
where use has been made of the fact that entropy of radiation on the current horizon is given by $S_\text{rad}=\frac{4\pi}{3}H_0^{-3}\frac{2\pi^2g_*(T_0)T_0^3}{45}\simeq 10^{90}$. This result ensures that the total entropy of the universe is so large that $\Omega = 1$ throughout the early epochs of cosmic evolution. The flatness problem concerns understanding why the (classical) starting conditions have led to a universe that was so close to spatial flatness (vanishing spatial curvature). It turns out that the ``natural value'' of the entropy is around unity at the Planck temperature when the horizon is around the Planck length. It grows to a large value at later times. 

On the other hand, as we found in (\ref{wave3a}), the natural size of the universe is not proportional to the Planck length, but it is 30 orders of magnitude greater than the Planck length rather than being proportional to it. Also, as was already seen in (\ref{abb1}), the entropy of radiation is set by the CC or, equivalently, by the degrees of freedom on the holographic screen
\begin{equation}
    \label{En3}
    S_\text{rad}=\frac{8\pi^\frac{1}{4}}{9}S_{dS}^\frac{3}{4}\simeq 10^{90}.
\end{equation}
The use of this radiation entropy in  (\ref{En2}) gives the size of the deviation from the flat spatial universe
\begin{equation}
    \label{En2a}
    \frac{|1-\Omega|}{\Omega}\Big|_{t=t_P}\simeq\frac{|k|}{\sqrt{S_{dS}}}\simeq 10^{-60}
\end{equation}
which is tiny enough to ensure that the large entropy of the cosmic event horizon solves the  flatness problem.

It is worth noting that the value of the radiation entropy (\ref{En3}) is precisely what is needed for solving the horizon problem. To see this, 
%The present horizon scale, $ct_0\simeq 10^{28}~ cm$, is at least as big as the current homogeneous and isotropic domain of the universe. By the ratio of the corresponding scale factors, $a_0/a_i$,this domain's initial size was smaller. We may reasonably infer that the homogeneous and isotropic region from which our universe formed at $t=t_i$ was bigger than $l_i$, assuming that inhomogeneity cannot be eliminated by the universe's expansion. Thus, $l_i\simeq t_0a_i/a_0$ \cite{mukhanov_2005}. This scale may easily be compared to the length $l_c\simeq t_i$ of a causal region
%\begin{equation}
%%    \label{wave8}
 %   \frac{l_i}{l_c}\simeq \frac{t_0}{t_p}\frac{a(t_P)}{a(t_0)}.
%\end{equation}
let us determine the spatial size of the visible universe during the Planck epoch. Recalling that the current size is about $l_{H,0}(t_0)\simeq 4.3\times10^{26}~{\rm m}$,  the size of this region at Planck time must have been \citep{mukhanov_2005}
\begin{equation}
    \label{new8}
    l_{H,0}(t_P)=\frac{a(t_P)}{a(t_0)} l_{H,0}(t_0)\simeq 10^{30}L_P,
\end{equation}
in accordance with the hot big bang theory. Let us first see how the above equation is interpreted in the standard big bang cosmology: The size of the universe at the Planck time ``must be'' the Planck length. Then, according to hot big bang cosmology, the observable universe at the time must have been 30 orders of magnitude larger than the natural size of a  causally-connected region. In other words, in the observable universe, the total number of causally unconnected regions at the start of the classical evolution must be about $10^{90}$. On the other hand, as we showed already in (\ref{wave3a}), the size of a causally-connected region after Planck times is equal to the size obtained in (\ref{new8}),  { see also Fig.\ref{fig2-3} (b)}. 
In other words, it is incorrect to believe that the initial size of the universe was around the Planck length, and our calculations demonstrate that the initial size of the universe (when classical evolution began) was far greater than the Planck length.

Needless to say, the CC (the vacuum energy) does not dilute during the expansion. The matter and radiation densities, on the other hand, dilute in time and give cause this way the cosmic coincidence problem. Indeed, the universe starts with radiation domination,  enters the matter era afterward, and finally (only recently around $z=1.3$) gets CC-dominated. If this CC domination  had occurred at an earlier epoch, the evolution of the universe would have been very different, and we would not be here to study it. 
As there is no clear fundamental physics explanation for why vacuum domination has  occurred only recently, many studies \citep{Garriga:1999bf,Garriga:1999hu,Bludman:1999gi} have concluded that some form of anthropic principle must have been in action. %This approach proposes a set of universes with varying vacuum energy values, the vast majority of which do not support the origin of life. In this regard, cosmic coincidence may be explained by arguing that the presence of intelligent life  selects only those vacuum energy density values which are comparable to those found in the supernova type Ia data. 
To see what the quantum cosmology approach tells about this, one can start with the  approximations (\ref{wave1aa}) and (\ref{wave6}) which lead to the Friedmann equation
\begin{equation}
    H^2=\frac{8\pi G}{3}\rho_\text{rad}+\frac{\Lambda}{3},
\end{equation}
with the radiation energy density 
\begin{equation}\label{new121}
    \rho_\text{rad}=\frac{S_{dS}}{6\pi^4}\frac{1}{a^4}.
\end{equation}
Now, using the radiation entropy in  (\ref{En3}) the radiation-to-CC ratio is found to be 
\begin{equation}
\label{ratio-r-2-cc}
    \frac{\Omega^\text{(rad)}}{\Omega^{(\Lambda)}}\simeq s^{\frac{4}{3}}L_P^4S_{dS}.
\end{equation}
Given that, at the present epoch, the radiation entropy density is $s=S_\text{rad}/a^3=2.8\times 10^3~{\rm cm}^{-3}$, one finds 
\begin{equation}
    \frac{\Omega^{(\text{rad})}_0}{\Omega_0^{(\Lambda)}}\simeq 10^{-5},
\end{equation}
as the value of the ratio  (\ref{RRR}) today. This estimate can be taken as an explanation of cosmic coincidence by the quantum cosmological approach. In other words, as equation (\ref{new121}) shows,
\begin{equation}
  \rho_0a_0^4\simeq S_{dS}\simeq \frac{1}{\Lambda L_P^2},   
\end{equation}
which means that the value of the radiation energy density at the present epoch is tied to the CC. %As a result, the ratio of CC and radiation densities is of the order of unity.

\section{Conclusions}

The genesis of the universe and the origin of the large-scale structure are two of the most fundamental and profound questions. To be more specific, there have been two critical questions in the history of modern cosmology: {\it (1)} Why is the universe on large scales nearly isotropic and homogeneous? {\it (2)} Why is the spatial curvature of the universe nearly zero (or rigorously zero)? The primary objective of inflationary models has been to give a dynamical explanation for these (and magnetic monopole) problems. And their explanation has been like when the potential energy of the inflaton field dominates the stress-energy distribution of a sufficiently large, (nearly) spatially-homogeneous region then the region experiences exponential expansion that dramatically expands it in size, isotropizes it and minimizes its spatial curvature. Inflation can also lead to a straightforward mechanism for structure formation.

It is nearly common folklore that inflation is regarded as the mere solution to the cosmological questions above. It does not, however, come without alternatives. Indeed, as was previously noted in \citep{Durrer:1995mz}, an oscillatory universe model can resolve the homogeneity problem. The matter bounce scenario \citep{Finelli:2001sr}, on the other hand, provides a viable  paradigm to have structure formation. String gas cosmology \citep{Brandenberger:2011et} is yet another alternative that explains the origin of structure in the universe as an alternative to inflation. Also important is that  a suitably large classical universe could emerge from Bohmian quantum cosmology without requiring a classical inflationary phase \citep{Pinto-Neto:2009kyd}. Notably, complex Hamilton-Jacobi formalism of quantum cosmology reveals the same possibility  \citep{Fathi:2016lws,Fathi:2017pjm}.

 In quantum gravity (and quantum cosmology), getting a large classical universe without an inflationary phase in the expanding classical era turns out to be hard because of the assumption that the typical linear size of the universe after leaving the quantum regime should remain around the Planck length. Consequently, the decelerated classical expansion after the quantum regime is insufficient to enlarge the universe in cosmic time $t$. More precisely, suppose that instead of holographic regularization of (\ref{2-34}), we use a usual regularization, for example, zeta function regularization \citep{Jalalzadeh:2022bgz}. Then, the contribution of the vacuum energy would be small, and consequently, the  scale factor at the Planck epoch would be proportional to the Planck length \citep{Jalalzadeh:2022bgz}. On the contrary, as we show in this paper, holographic regularization adds finite but enormous vacuum energy to the WDW equation, resulting in 30 orders of magnitude larger universe in the Planck epoch.

In the present work, we have also studied a holographic approach to the cosmic coincidence problem considering a minisuperspace under radiation domination. We have shown that the electromagnetic vacuum energy, regularized with a local UV cutoff and tied to global IR cutoff via holography, sets the state of the WDW equation in a closed, homogeneous, and isotropic universe. It leads to both expanding (and contracting) solutions with successful predictions for the scale factor for both Planckian and later times. The model successfully predicts the coincidence of the cosmological constant and radiation contributions today. The model can be extended by including matter besides the radiation and resolving their respective evolutions in time. The model can also be extended by considering an emergent gravity approach (like the symmergent gravity noted before) in which vacuum energy and boson masses could be under better control to have a more predictive model by exploiting the holography again.

\section*{Acknowledgements}
S.J. acknowledges financial support from the National Council for Scientific and Technological Development--CNPq, Grant no. 308131/2022-3. D.D. thanks to Ozan Sarg{\i}n for discussions. 
\bibliographystyle{elsarticle-num}

\bibliography{main}

\end{document}